%% file: conference_101719.tex
\documentclass[conference]{IEEEtran}
\IEEEoverridecommandlockouts
\usepackage{cite}
\usepackage{amsmath,amssymb,amsfonts}
\usepackage{algorithmic}
\usepackage{graphicx}
\usepackage{textcomp}
\usepackage{xcolor}
\usepackage[nolist]{acronym}
\usepackage{paralist}
\usepackage{booktabs}
\def\BibTeX{{\rm B\kern-.05em{\sc i\kern-.025em b}\kern-.08em
    T\kern-.1667em\lower.7ex\hbox{E}\kern-.125emX}}

\begin{document}


\input{acronyms}

\bstctlcite{IEEEexample:BSTcontrol}

\title{On Process Awareness in Detecting Multi-stage Cyberattacks in Smart Grids}

\author{
\IEEEauthorblockN{%
Ömer Sen\IEEEauthorrefmark{1}\IEEEauthorrefmark{2},
Yanico Aust\IEEEauthorrefmark{1},
Simon Glomb\IEEEauthorrefmark{1},
Andreas Ulbig\IEEEauthorrefmark{1}\IEEEauthorrefmark{2},
}

\IEEEauthorblockA{%
\IEEEauthorrefmark{1}\textit{Digital Energy, Fraunhofer FIT} Aachen, Germany\\
Email: \{oemer.sen, andreas.ulbig\}@fit.fraunhofer.de}
\IEEEauthorblockA{%
\IEEEauthorrefmark{2}\textit{IAEW, RWTH Aachen Univserity,} Aachen, Germany\\
Email: \{o.sen, a.ulbig\}@iaew.rwth-aachen.de | \{yanico.aust, simon.glomb\}@rwth-aachen.de} 
}

\IEEEoverridecommandlockouts


\maketitle

\IEEEpubidadjcol

\begin{abstract}
This study delves into the role of process awareness in enhancing intrusion detection within Smart Grids, considering the increasing fusion of ICT in power systems and the associated emerging threats. The research harnesses a co-simulation environment, encapsulating IT, OT, and ET layers, to model multi-stage cyberattacks and evaluate machine learning-based IDS strategies. The key observation is that process-aware IDS demonstrate superior detection capabilities, especially in scenarios closely tied to operational processes, as opposed to IT-only IDS. This improvement is notable in distinguishing complex cyber threats from regular IT activities. The findings underscore the significance of further developing sophisticated IDS benchmarks and digital twin datasets in Smart Grid environments, paving the way for more resilient cybersecurity infrastructures.
\end{abstract}


\input{chapter1}
\input{chapter2}
\input{chapter3}
\input{chapter4}
\input{chapter5}
\vspace{-0.5em}

\section*{Acknowledgment}
\begin{minipage}{0.65\columnwidth}%
This work has received funding from the BMBF under project no. 03SF0694A (Beautiful).
\end{minipage}
\hspace{0.02\columnwidth}
\begin{minipage}{0.23\columnwidth}%
	\includegraphics[width=\textwidth]{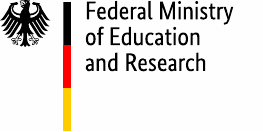}
\end{minipage}
\vspace{-0.5em}

\bibliographystyle{IEEEtran}
\bibliography{conference_101719}

\end{document}

%% file: acronyms.tex
\begin{acronym}
\acro{sg}[SG]{smart grid}
\acroplural{sg}[SGs]{smart grids}
\acro{der}[DER]{distributed energy resource}
\acroplural{der}[DERs]{distributed energy resources}
\acro{ict}[ICT]{information and communication technology}
\acro{fdi}[FDI]{false data injection}
\acro{scada}[SCADA]{Supervisory Control and Data Acquisition}
\acro{mtu}[MTU]{Master Terminal Unit}
\acroplural{mtu}[MTUs]{Master Terminal Units}
\acro{hmi}[HMI]{Human Machine Interface}
\acro{plc}[PLC]{Programmable Logic Controller}
\acro{dmz}[DMZ]{Demilitarized Zone}
\acroplural{plc}[PLCs]{Programmable Logic Controllers}
\acro{ied}[IED]{Intelligent Electronic Device}
\acroplural{ied}[IEDs]{Intelligent Electronic Devices}
\acro{rtu}[RTU]{Remote Terminal Unit}
\acro{et}[ET]{Electrical Technology}
\acroplural{rtu}[RTUs]{Remote Terminal Units}
\acro{iec104}[IEC-104]{IEC 60870-5-104}
\acro{apdu}[APDU]{Application Protocol Data Unit}
\acro{apci}[APCI]{Application Protocol Control Information}
\acro{asdu}[ASDU]{Application Service Data Unit}
\acro{io}[IO]{information object}
\acroplural{io}[IOs]{information objects}
\acro{cot}[COT]{cause of transmission}
\acro{mitm}[MITM]{Man-in-the-Middle}
\acro{fdi}[FDI]{False Data Injection}
\acro{ids}[IDS]{intrusion detection system}
\acroplural{ids}[IDSs]{intrusion detection systems}
\acro{siem}[SIEM]{Security Information and Event Management}
\acro{mv}[MV]{medium voltage}
\acro{lv}[LV]{low voltage}
\acro{cdss}[CDSS]{controllable distribution secondary substation}
\acro{bss}[BSS]{battery storage system}
\acroplural{bss}[BSSs]{battery storage systems}
\acro{pv}[PV]{photovoltaic inverter}
\acro{mp}[MP]{measuring point}
\acroplural{mp}[MPs]{measuring points}
\acro{dsc}[DSC]{Dummy SCADA Client}
\acro{fcli}[FCLI]{Fronius CL inverter}
\acro{fipi}[FIPI]{Fronius IG+ inverter}
\acro{sii}[SII]{Sunny Island inverter}
\acro{tls}[TLS]{Transport Layer Security}
\acro{actcon}[ActCon]{Activation Confirmation}
\acro{actterm}[ActTerm]{Activation Termination}
\acro{rtt}[RTT]{Round Trip Time}
\acro{c2}[C2]{Command and Control}
\acro{dst}[DST]{Dempster Shafer Theory}
\acro{ec}[EC]{Event Correlator}
\acro{sc}[SC]{Strategy Correlator}
\acro{ioc}[IoC]{Indicator of Compromise}
\acroplural{ioc}[IoCs]{Indicators of Compromise}
\acro{ot}[OT]{Operational Technology}
\acro{it}[IT]{Information Technology}
\acro{aucmorf}[AUC-MORF]{Area Under the Most Relevant First Perturbation Curve}
\acro{aupc}[AUPC]{Area Under Perturbation Curve}
\acro{ics}[ICS]{Industrial Control System}
\acroplural{ics}[ICSs]{Industrial Control Systems}
\acro{pera}[PERA]{Purdue Enterprise Reference Architecture}
\acro{dpi}[DPI]{Deep Package Inspection}

\acro{iec104}[IEC104]{IEC 60870-5-104}
\acro{rtu}[RTU]{Remote Terminal Unit}
\acro{mtu}[MTU]{Master Terminal Unit}
\acro{dos}[DoS]{Denial of Service}
\acro{mitm}[MITM]{Man-in-the-Middle}
\acro{arp}[ARP]{Address Resolution Protocol}
\acro{tcp}[TCP]{Transmission Control Protocol}
\acro{pcap}[PCAP]{Packet Capture}
\acro{ids}[IDS]{Intrusion Detection System}
\acro{ics}[ICS]{Industrial Control Systems}
\acro{sg}[SG]{Smart Grids}
\acro{der}[DER]{Distributed Energy Resources}
\acro{ot}[OT]{Operational Technology}
\acro{scada}[SCADA]{Supervisory Control and Data Acquisition}
\acro{coa}[COA]{common address of ASDU}
\acro{coas}[COA]{common addresses of ASDU}
\acro{ict}[ICT]{Information and Communication Technology}
\acro{it}[IT]{Information Technology}
\acro{et}[ET]{Energy Technology}
\acro{dmz}[DMZ]{Demilitarized Zone}
\acro{ied}[IED]{Intelligent Electronic Device}
\acro{plc}[PLC]{Programmable Logic Controller}
\acro{hmi}[HMI]{Human Machine Interface}
\acro{cps}[CPS]{Cyber Physical System}
\acro{dso}[DSO]{Distribution System Operator}
\acro{ova}[OVA]{One-Vs-All}
\end{acronym}

%% file: chapter1.tex
\section{Introduction}
The transformation of traditional power grids into \acp{sg} has brought significant advancements and challenges, particularly in the realm of cybersecurity due to enhanced connectivity~\cite{pan2015developing}. The integration of \ac{ict} components, while beneficial, has exposed these grids to cyber threats, making robust cybersecurity measures essential.

One of the primary tools in combating these threats is the use of \acp{ids}, which have traditionally relied heavily on \ac{ict} data, potentially overlooking critical underlying processes. The development of process-aware \acp{ids} represents a significant enhancement in this field, focusing on domain-specific knowledge to detect anomalies accurately within the unique operational context of \acp{sg}~\cite{shun2008network}. These systems are designed to recognize abnormal behavior patterns and cybersecurity threats effectively.

Despite advancements in \acp{ids}, challenges remain, such as improving detection accuracy, reducing false positives, and coping with novel attack vectors. Machine learning techniques have emerged as a promising solution to enhance the capabilities of \acp{ids} through their ability to learn and adapt to new threats~\cite{liu2019machine}. However, the application of these methods still struggles with high false positive rates and unclear alert mechanisms, which complicates their deployment in critical infrastructure~\cite{samrin2017review}. 

The need for an effective anomaly detection system is paramount, particularly one that offers high precision and clear, explainable alerts. This is crucial for reducing the effort required to process alerts and for improving the speed and accuracy of responses to potential threats.

In environments like \ac{ics} and \acp{sg}, detecting anomalies in process data is critical. Techniques employed in machine learning-driven \acp{ids} include allowlisting and blocklisting, which help identify deviations from normal operations and focus on predefined abnormal behaviors. These capabilities are essential for detecting various issues like cyberattacks, system errors, or configuration mistakes, which may manifest as protocol anomalies or communication disruptions.

This paper addresses several challenges:
\begin{itemize}
\item Leveraging domain knowledge inherent to electrical grid operations for detecting cyber threats and enhancing \ac{ids} capabilities.
\item Integrating indicators from diverse domains such as \ac{it}, \ac{ot}, and \ac{et} to improve attack detection in \acp{sg}.
\item Employing process-aware detection techniques to understand the impact of different domains on attack detection.
\end{itemize}

Our research explores the effectiveness of process-awareness in identifying complex, multi-stage cyberattacks in \acp{sg}, combining \ac{sg} knowledge with traditional \ac{ict} intrusion detection methods. We utilize a detailed simulation environment that reflects all aspects of \ac{sg} operations, including \ac{it}, \ac{ot}, and \ac{et} components. A layered approach is employed where sensors monitor the grid for anomalies and relay information to a central system that integrates domain knowledge for enhanced process awareness. This system analyzes both \ac{ict} traffic and process data from \ac{scada} systems, correlating different data types to form a comprehensive understanding of potential cyber incidents.

Our contributions include investigating the effectiveness of process-aware \acp{ids}, employing comprehensive simulation environments, and implementing a layered approach to enhance grid security through improved anomaly detection.


%% file: chapter2.tex
\section{Process Awareness in IDS}\label{background:ov}
In this section, we develop an understanding of \ac{ids}, elaborate on the characteristics of process awareness, and analyze machine learning-based methods for anomaly detection.

\begin{figure}\textbf{}
    \centering
    \includegraphics[width=\columnwidth]{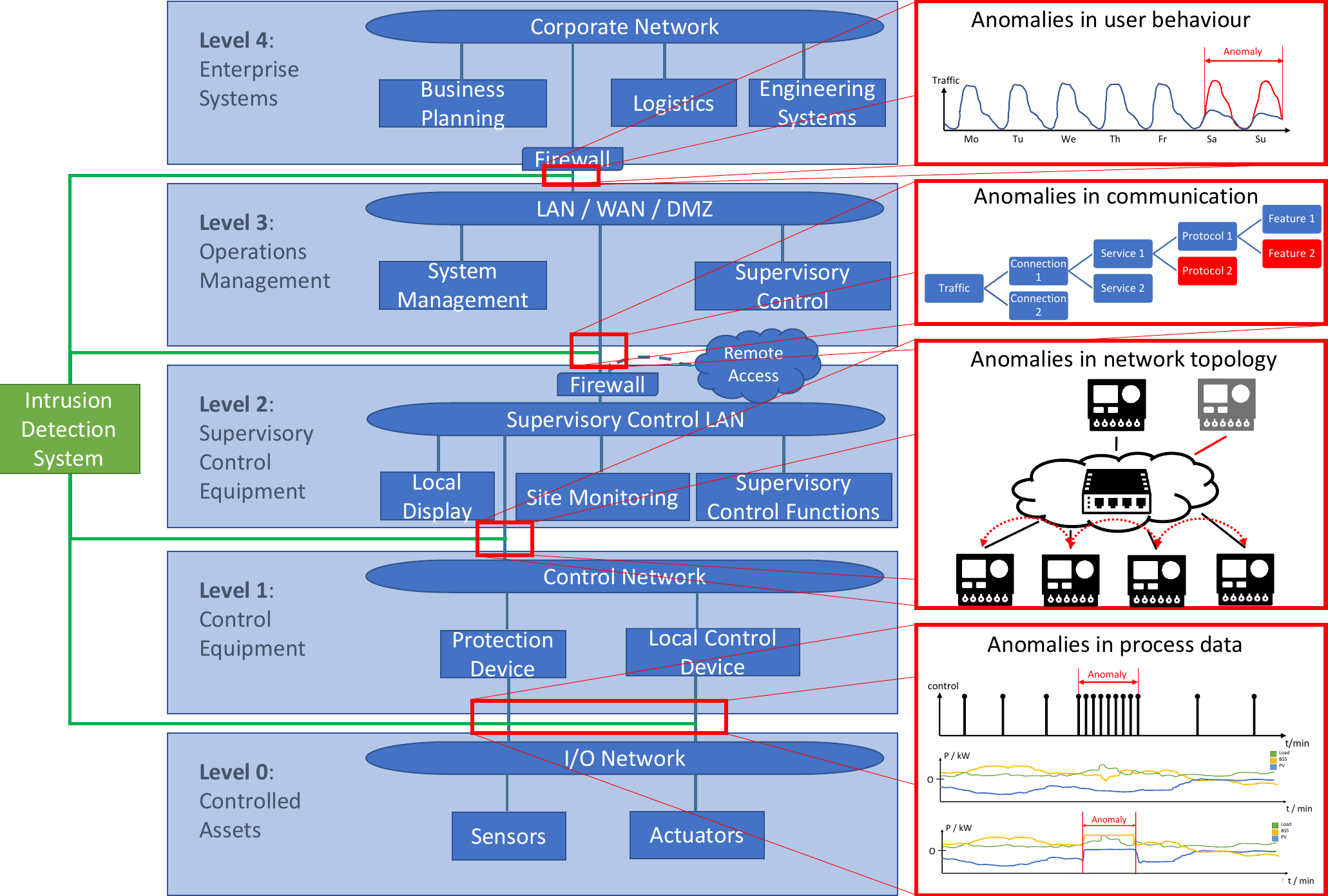}
    \caption{Illustration of \ac{ids} integration into \ac{sg} industrial control architecture and anomalies in the environment}
    \label{fig:ids_in_ics}
\end{figure}

\subsection{Intrusion Detection Systems}\label{background:ids}
An \ac{ids} is designed to monitor network or system activities to detect cyber threats, such as unauthorized access or misuse, through both software and hardware solutions. \acp{ids} are vital in \ac{ics} environments, particularly \ac{sg}, due to their capability to monitor device logs, network traffic, and process data, identifying suspicious activities~\cite{hu2009simple}. 
Host-based \acp{ids} focus on internal device activities, while network-based \acp{ids} analyze external network traffic. Process-aware \acp{ids} in \ac{sg} are crucial for detecting anomalies by analyzing process data, including measurements or control commands.

Machine learning methods in \acp{ids} include allowlisting and blocklisting, which identify deviations from normal operations or recognize predefined abnormal behaviors, respectively. These methods are essential for detecting various anomalies, such as protocol errors and misconfigurations. Explainability in these systems helps in understanding the basis of alerts and maintaining system trust~\cite{islam2020towards}.

Machine learning in \acp{ids} involves significant preprocessing steps like data cleaning, transformation, and reduction. Ensemble learning, particularly stacking methods, is employed to improve classification performance by combining multiple learning algorithms. These models are evaluated based on detection accuracy, runtime efficiency, and explainability, with a focus on minimizing false positives and enhancing system trust~\cite{liu2019machine}. 

\subsection{Process Awareness}\label{background:processawanress}
Process-awareness in \ac{ids} utilizes domain-specific knowledge to enhance anomaly detection, particularly in \ac{sg} environments (cf. Figure~\ref{fig:ids_in_ics}). This approach leverages confirmed communication patterns and network structures to detect anomalies, focusing on unknown devices, new connections, or altered communication functions. Anomalies can arise from cyberattacks or system faults, manifesting as protocol errors or delayed messages. Network-based interactions and the traffic’s special information aid in detecting anomalies and assessing the legitimacy of system components.

The Purdue Model of \ac{pera} categorizes \ac{ics} architecture into operational (OT) and informational (IT) zones, further divided into six levels from physical hardware to enterprise network integration~\cite{williams1996overview}. This framework helps in selecting datasets for benchmarking \acp{ids} across different \ac{ics} levels, focusing on anomalies that affect both physical components and control systems.

Datasets that mirror real-world scenarios at various \ac{ics} levels are crucial for evaluating \acp{ids} effectiveness in detecting and classifying network-based threats. This comprehensive approach ensures \acp{ids} are tested against diverse attack types, enhancing their robustness and applicability in \ac{ics} environments.

\subsection{Related Work}\label{background:related_work}
The research landscape has been actively exploring the development of benchmark environments to evaluate the performance of various machine learning algorithms in detecting anomalies and intrusions within \ac{ics}. Approaches like the Penn Machine Learning Benchmark~\cite{olson2017pmlb} and the Scientific Machine Learning Benchmark suite~\cite{thiyagalingam2022scientific} provide vital resources for testing and comparing algorithm performance. Researchers have also emphasized the creation of datasets from real \ac{ics} environments, such as the Cyber-kit datasets~\cite{mubarak2021ics} and the Numenta anomaly benchmark~\cite{lavin2015evaluating}, to evaluate unsupervised anomaly detection techniques or \acp{ids} in \ac{ics}. Specific focus has been given to datasets related to \acp{sg} to assess the performance of machine learning algorithms in detecting anomalies in these systems~\cite{bernieri2019evaluation, liyakkathali2020validating, mohammadpourfard2019benchmark}. However, there exists a need for more specialized and comprehensive approaches, particularly in considering the unique characteristics and constraints of \ac{ics} and \acp{sg}, such as explainability of algorithms and statistical confidence of findings~\cite{japkowicz2006question, tufan2021anomaly}.

In addition, research works have investigated process-aware \acp{ids} in \ac{ics}, focusing on evaluating the criticality of \ac{ics} devices and identifying potential adversary traces~\cite{15_cook2017industrial, 16_escudero2018process}. Advanced approaches include the replication of program states in digital twins~\cite{9_eckhart2018specification}, cyber-attack classification~\cite{18_mohan2020distributed}, and modeling \ac{ics}/\ac{scada} communication using probabilistic automata~\cite{19_matouvsek2021efficient, 20_almseidin2019fuzzy}. Anomaly detection methods for \ac{iec104} have also been explored using multivariate access control and outlier detection approaches~\cite{grammatikis2020anomaly, burgetova2021anomaly, anwar2021comparison}. However, these proposed approaches require additional analytical resources for their functionality, such as infrastructure specifications, attack target understanding, statistical data, or technical specifications~\cite{scheben2017status, dang2021improving, holzinger2020measuring}. Process awareness in \ac{ids} has not been systematically investigated, particularly in terms of its specific contribution to the general detection of cyberattacks and the added value it provides across different domain spaces of \ac{ics}.

This paper aims to delve into the realm of process awareness in \ac{ids}, particularly focusing on how domain knowledge from the operation of electrical grids can be leveraged to enhance these systems. A key aspect of our investigation is the identification of significant domain knowledge inherent to grid operation that could be crucial for detecting cyberattacks. This includes an in-depth analysis of the plausibility of process data, a critical factor in discerning legitimate operations from potential security breaches. We also explore the integration of relevant indicators from diverse domains, specifically \ac{it}, \ac{ot}, and \ac{et}. This involves identifying domain-specific attributes of the indicators with regard to detecting multi-stage cyber attacks impacting several domains in the \ac{ics} layer hierarchy.

%% file: chapter3.tex
\begin{figure}[htbp]
    \centering
    \includegraphics[width=\columnwidth]{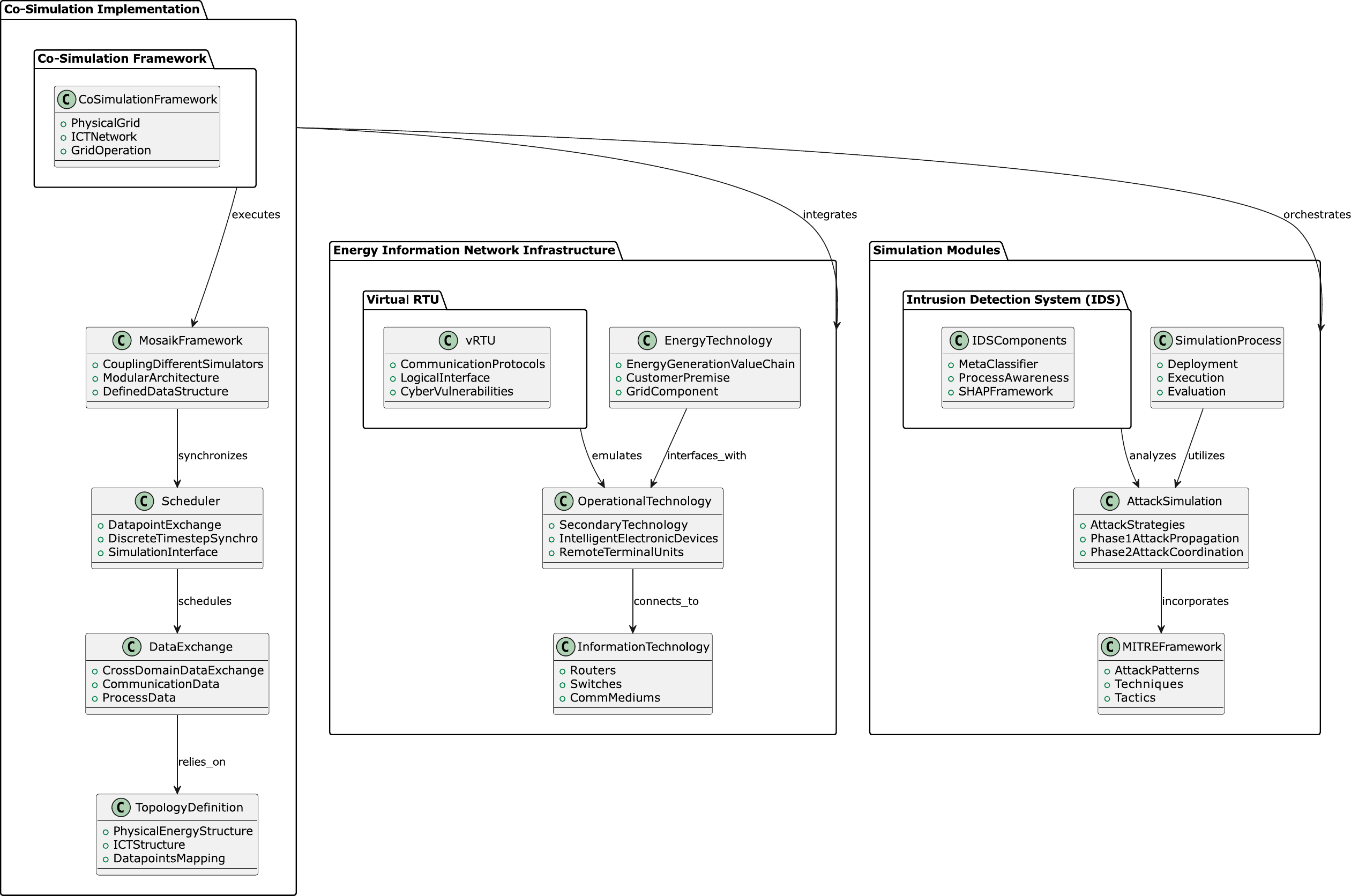}
    \caption{Overview of the class diagrams of the co-simulation for the investigation of process-aware \ac{ids}}
    \label{component_diag}
\end{figure}

\section{Investigation Framework}\label{sec:methodology}
This section outlines our research framework developed to investigate how process-awareness can enhance \ac{ids}, especially within \ac{sg} environments.

\subsection{Overview}\label{sec:methodology_overview}
The framework emphasizes the integration of domain knowledge from both \ac{it} and \ac{ot} sectors, aiming to detect and differentiate cyber threats from genuine operational activities. By analyzing process data, the approach not only distinguishes between normal and anomalous activities but also considers that system faults may not always indicate cyberattacks. 

It involves a detailed correlation and combination of indicators across \ac{it} and \ac{ot} domains, enhancing the effectiveness and proactivity of \ac{ids} in managing grid security. The visual representation in Figure~\ref{component_diag} highlights the dynamic interplay between energy technologies and operational technologies within a co-simulation environment. This illustration depicts the complex interactions essential for realistic security investigations and the framework used to evaluate the plausibility of detected activities.

Reproducibility is critical for scientific validation and requires well-documented methodologies encompassing datasets, data preprocessing, model training, and evaluation~\cite{tatman2018practical, uetz2021reproducible}. This study utilizes domain-specific, balanced datasets based on the Purdue Model~\cite{williams1996overview}. Preprocessing includes data cleaning, transformation, reduction, and labeling~\cite{davis2011data}. We emphasize the use of ensemble learning techniques to enhance detection accuracy and employ robust evaluation metrics like detection rate, runtime efficiency, and explainability~\cite{zhou2012ensemble, liu2019machine}.

\subsection{Investigation Environment}\label{subsec:methodology-investigation}
Our work focuses on developing a comprehensive co-simulation framework for energy information networks, merging \ac{dso} infrastructure across \ac{et}, \ac{ot}, and \ac{it} domains. The \ac{et} domain manages essential elements like switches and transformers for efficient electricity distribution, while the \ac{ot} domain features \acp{ied} connecting primary units to support crucial functions like control and measurement, primarily communicating via \ac{iec104} protocols to the \ac{scada} system. The \ac{it} domain handles remote maintenance and external communications, with the \ac{scada} system playing a critical role, linking the \ac{ot} network with the office network via a \ac{dmz}.

The simulation emphasizes a component-based, modular approach, accommodating diverse network characteristics. It is designed to provide a realistic depiction of component behaviors and communication within \ac{it} and \ac{ot} domains, crucial for representing \ac{ict} failures and cyberattacks. This modular simulation supports varied investigative scenarios, enhancing the resilience of energy information networks.

The co-simulation process comprises three phases: Deployment, involving network topology setup and environment establishment; Execution, managing synchronized data transfer; and Evaluation, focusing on result analysis, especially the impact of cyberattacks.

Utilizing the Mosaik framework, our co-simulation facilitates central initialization, scheduling, and data exchange among simulators~\cite{schutte2011mosaik}, supporting large-scale scenarios and integration with real-time labs and software-in-the-loop systems. The "rettij" simulator integrates into this environment to emulate realistic \ac{ict} communication networks, using real network stacks and standard Linux tools to create complex network topologies~\cite{niehaus2022modern}.

For modeling attacks, we apply the MITRE ATT\&CK Framework, which categorizes attack modalities and patterns. Our cyberattack simulation stages include Initial Access, Execution, Privilege Escalation, Credential Access, Lateral Movement, Collection, Command and Control, and Impact, each reflecting specific attack objectives and techniques.

This co-simulation framework provides a robust platform for scrutinizing and enhancing the security of energy information networks, focusing on genuine scenarios and comprehensive vulnerability evaluations.

\subsection{Multistage Attack Implementation}\label{subsec:mutlistage_implementation}
The implementation of multistage cyberattacks across the main domains \ac{et}, \ac{ot}, and \ac{it} is detailed in this section. These attacks are essential for a comprehensive coverage of the network infrastructure and are reused in various multistage scenarios due to their efficacy.

The setup includes an additional node simulating the attacker, equipped with a suite of adversarial tools, allowing for prolonged surveillance and deep insight into network operations such as \acp{rtu}, \acp{mtu}, and switches. Initial network access is assumed, setting the stage for a series of coordinated attacks.

One of the primary techniques employed is \ac{arp} Spoofing/Poisoning. The attacker manipulates the ARP protocol to redirect legitimate IP addresses to their machine, effectively camouflaging malicious activities. This facilitates \ac{mitm} attacks, where traffic between specific devices is rerouted through the attacker, enabling packet sniffing and modification.

Network connection enumeration is another critical strategy, utilizing subnet scanning and port scanning to identify and exploit vulnerabilities within the network.

Specific attacks include:
\begin{inparaenum}[(i)]
\item  A \ac{dos} attack targeting the network's \acs{mtu}, using a \acs{tcp} 3-way handshake to disrupt service to 26 \acs{rtu}, maximizing the impact on network operations.
\item  The \ac{iec104} Value Modification attack, initiated once a \acs{mitm} position is established, involves intercepting and altering \ac{iec104} protocol traffic. This attack can be precise, targeting specific system commands to cause disruption or confusion.
\item  The Replay Attack, which captures legitimate traffic patterns between the \acs{mtu} and \acp{rtu}, later replaying altered packets to simulate normal activities, thus masking the intrusion.
\end{inparaenum}

These simulated multistage attacks demonstrate complex strategies to exploit network vulnerabilities, emphasizing the need for robust security measures in power grids.

\subsection{IDS Implementation}\label{subsec:methodology-ids}
This section elaborates on the classifier used in our evaluations, employing the \ac{ova} technique for multi-class classification~\cite{rifkin2004defense}. Here, \textit{N} binary classifiers are trained for \textit{N} classes, with each classifier differentiating its class from all others. Predictions from these classifiers are integrated by a meta-classifier, which selects the most confident output, enhancing multi-stage cyberattack detection.

The meta-classifier employs a windowing technique to analyze segments of data for potential attack patterns, utilizing the softmax function to convert classifier outputs into a probability distribution. This nuanced approach allows for detecting and sequencing multi-stage attacks, essential for recognizing deviations from expected attack patterns.

Real-time capabilities are crucial for \acp{ids}, hence a windowing mechanism is employed to analyze data within specific time frames, focusing on sequential detection of attack stages, which may include typical sequences like initial access followed by execution and persistence.

The SHAP framework is implemented to determine the impact of individual features on predictions, applied to each base classifier within our architecture~\cite{chen2020true}. This helps in understanding the decision-making process by elucidating how each classifier's outputs contribute to the final decision. If the combination process lacks transparency, SHAP can reassess the influence of each base classifier.

In our setup, data from ''unified2'' files is parsed into a DataFrame, preprocessed to remove redundant columns and encode categorical data, with labels binarized for use with the meta-classifier. The analysis involves supervised, unsupervised, and mixed learning techniques to detail feature contributions using the SHAP framework, visualized through bar and beeswarm plots.

Our methodology, combining a meta-classifier with the SHAP framework, provides a robust system for real-time detection of multi-stage cyberattacks. This holistic approach ensures precise and clear threat detection, leveraging varied data inputs to enhance intrusion detection accuracy and reliability in complex environments like smart grids.

Our \ac{ids} approach utilizes an ensemble stacked meta-learner approach, where different base learners are associated with different phases of the attack, while the overarching meta-learner is responsible for determining the overall attack classification. Particularly, we associate the MITRE ATT\&CK Matrix for \ac{ics} with the different base learners, where we traditionally distinguish between \ac{it} and \ac{ot} related phases. Table~\ref{table:simplified_mitre_attack_phases} describes the mapping of the MITRE ATT\&CK phases to the \ac{it} and \ac{ot} stages. 

\vspace{-1em} 

\begin{table}[ht!]
\centering
\caption{Simplified MITRE ATT\&CK Phases for IT and OT in ICS}
\label{table:simplified_mitre_attack_phases}
\begin{tabular}{|p{1.7cm}|p{2.9cm}|p{2.9cm}|}
\hline
\textbf{Phase} & \textbf{IT Attacks} & \textbf{OT Attacks} \\ \hline
Initial Access & Phishing, System exploits & Physical, Workstation exploits \\ \hline
Execution & Malware, Scripting & Through HMIs, Field devices \\ \hline
Persistence & Account, Registry keys & Firmware, Device replacement \\ \hline
Privilege Esc. & Exploits, Rootkits & Bypass security controls \\ \hline
Defense Evasion & Obfuscation, Log deletion & Logic tampering, State changes \\ \hline
Credential Access & Credential dumping, Sniffing & Credential compromise in controllers \\ \hline
Discovery & Service scanning, System discovery & Sniffing, ICS network scanning \\ \hline
Lateral Movement & Pass the Hash, Pivoting & Interconnected system compromise \\ \hline
Collection & Local and network data & Process system, Historian data \\ \hline
C2 & Port use, Encryption & Server comms, Tunneling \\ \hline
Exfiltration & Compression, Scheduled transfer & Over C2 channels, Replication \\ \hline
Impact & Data destruction, Disruption & Process manipulation, Sabotage \\ \hline
\end{tabular}
\end{table}

To provide process-awareness, we segregated the relevant fields into their fitting categories. The structured overview of the used fields is provided in Table~\ref{table:field_descriptions}, listing each indicator according to their operational significance, categorizing them into their specific domains (\ac{it}, \ac{ot}, and \ac{et}), and facilitating the efficient interpretation of dissected traffic data. This structured approach leads to a comprehensive interpretation of industrial protocol traffic within a cyber-physical environment for IDS purposes. Finally, process-awareness is incorporated into the \ac{ids} with the capability of \ac{dpi}, where fields related to \ac{it}, \ac{ot}, and \ac{et} in the industrial protocol traffic are dissected and interpreted accurately. We modified our \ac{ids}, such that it would provide seven different categorical types of results by providing each possible combination from \ac{it}, \ac{ot}, and \ac{et}. The \ac{ids} creates security events according to Table~\ref{table:simplified_mitre_attack_phases} and masks out only the required fields that fit in the corresponding category. Since our approach tests and evaluates all possible combinations, the best possible combination of categories is created and gives insights into the importance of process-awareness.

\vspace{-1em}

\begin{table}[ht!]
\centering
\caption{Fields of event describing indicators}
\begin{tabular}{|p{2.2cm}|p{4.2cm}|p{1.1cm}|}
\hline
\textbf{Field Name} & \textbf{Description}  & \textbf{Category} \\
\hline
timestamp & Event timestamp & Global \\
categorization & Event categorization & Global \\
priority & Event priority level  & Global\\
phase & Phase in MITRE ATT\&CK & Global\\
ttp & Tactics, techniques, procedures & Global \\
id & Unique event identifier & Global\\
IP Data & Various IP Layer protocol fields & IT\\
TCP Data & Various TCP Layer protocol fields & IT\\
rtt & Packet round-trip time & IT\\
frequency\_general & General event frequency & IT\\
frequency\_proto & Protocol-specific event frequency & IT \\
throughput & Network throughput & IT\\
iec104\_frame & IEC 104 frame format & OT\\
diff\_tx & TX difference between two packages&OT\\
diff\_rx &  RX difference between two packages&OT\\
iec104\_type\_id & IEC 104 type identification&OT\\
iec104\_oa & IEC 104 origin address&OT\\
iec104\_numix & IEC 104 number of objects&OT\\
iec104\_coa & IEC 104 common ASDU address&OT\\
iec104\_ioa & IEC 104 information object address&OT\\
iec104\_cot & IEC 104 cause of transmission&OT\\
iec104\_value\_sigma &Sigma between IEC 104 IO values&OT\\
iec104\_io\_value & IEC 104 protocol IOA value & ET\\
iec104\_control & IEC 104 control signals & ET\\
iec104\_status & IEC 104 status notification & ET\\
\hline
\end{tabular}
\label{table:field_descriptions}
\end{table}

%% file: chapter4.tex
\vspace{-1em}

\section{Investigation}\label{sec:result}
In this section, we present our findings on the effectiveness of process-aware \ac{ids}. The investigation involves a scenario to simulate a multi-stage cyberattack, enabling a detailed analysis of detection capabilities across different network layers.

\begin{figure}[htbp]
    \centering
    \includegraphics[width=\columnwidth]{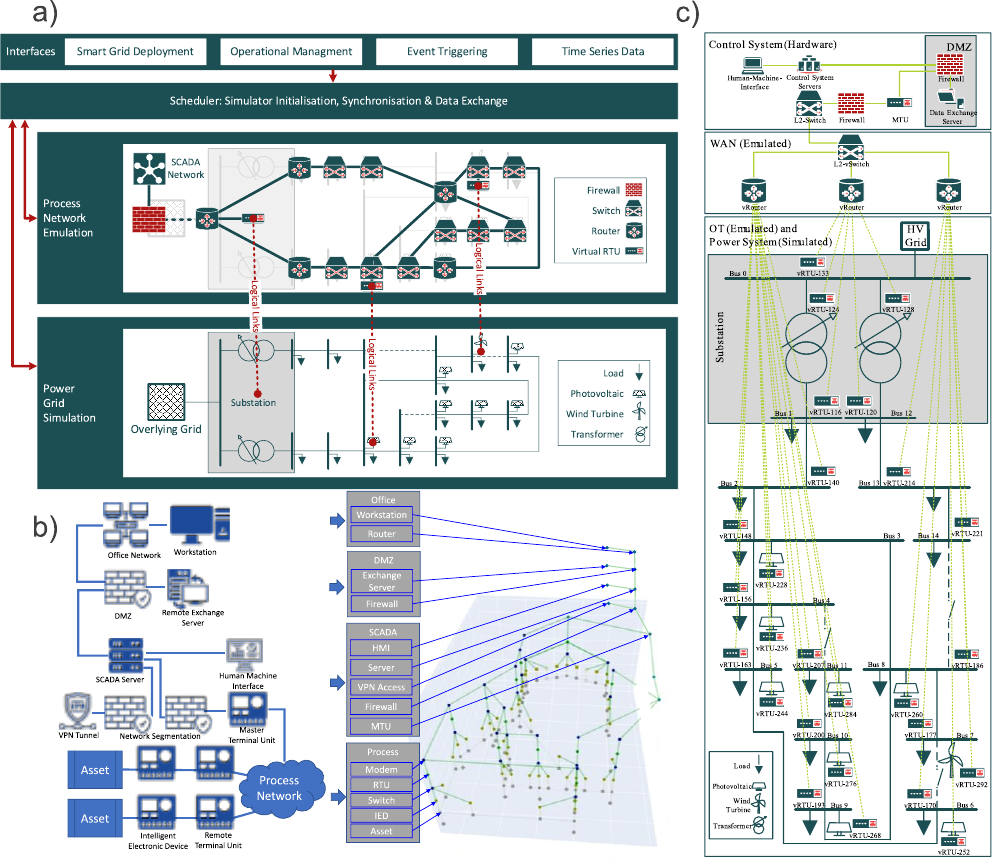}
    \caption{Overview of the cigre-based scenario simulated in the co-simulation for the study: a) the co-simulation architecture and the components involved, b) the infrastructure model of the scenario and c) the specific network model for the case}
    \label{fig:scenario_overview}
    \vspace{-1em}
\end{figure}

\subsection{Scenario}\label{sec:result_procedure}
In this study, we utilize the CIGRE grid model to simulate a distribution grid with integrated decentralized power generation~\cite{pandapowerCigre} (cf. Figure~\ref{fig:scenario_overview}). This scenario features an \ac{ot} communication infrastructure comprising routers, switches, firewalls, \ac{scada} systems, and \acp{rtu}, all interconnected to ensure control compliant with \ac{iec104} standards, facilitating the simulation of power flow and communication within a smart grid framework.

Our multi-stage cyberattack employs MITRE ATT\&CK recommended techniques in a sequenced manner. Initially, \ac{arp} spoofing between \acp{rtu} and an \ac{mtu} establishes a \ac{mitm} link, allowing packet manipulation including alteration and dropping. During the first 10\% of the simulation, \ac{tcp} RST flags are manipulated, followed by modifications to \ac{iec104} transmission causes and information object values between 20\% and 80\% of the simulation time.

The final 20\% features an SSH brute-force attack on a non-spoofed \ac{rtu}, carefully executed to avoid triggering alarms. Post-attack, an \ac{ids} processes PCAP files to detect anomalies, analyzing traffic and matching specific patterns to improve accuracy. This \ac{ids} categorizes events into seven groups corresponding to \ac{it}, \ac{ot}, and \ac{et} layers, each undergoing separate classification to ensure balanced representation of all attack techniques.

Additionally, the simulation can replicate normal operational scenarios where no attacks occur but \ac{ict} failures like communication link breakups are present. The \ac{ids} analysis is performed using PCAP files that record all network traffic, focusing on pattern matching and attack detection across the \ac{it}, \ac{ot}, and \ac{et} layers. Each network event is categorized and analyzed for its relevance to detection, with a category classification process for balanced representation.

\subsection{Case Studies}\label{sec:result_findings}
\begin{figure}[htbp]
    \centering
    \includegraphics[width=\columnwidth]{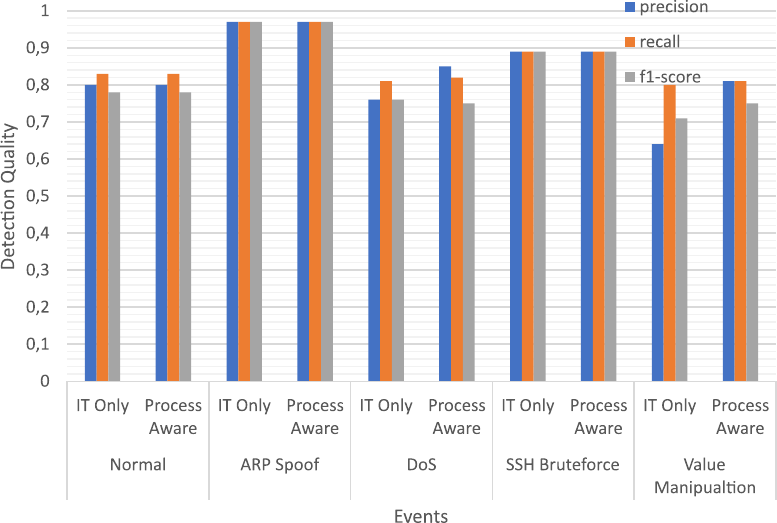}
    \caption{Comparison between the \ac{it}-only and process-aware \ac{ids} in various scenarios}
    \label{fig:estimation_plot}
    \vspace{-1em}
\end{figure}

\begin{figure}[htbp]
    \centering
    \includegraphics[width=\columnwidth]{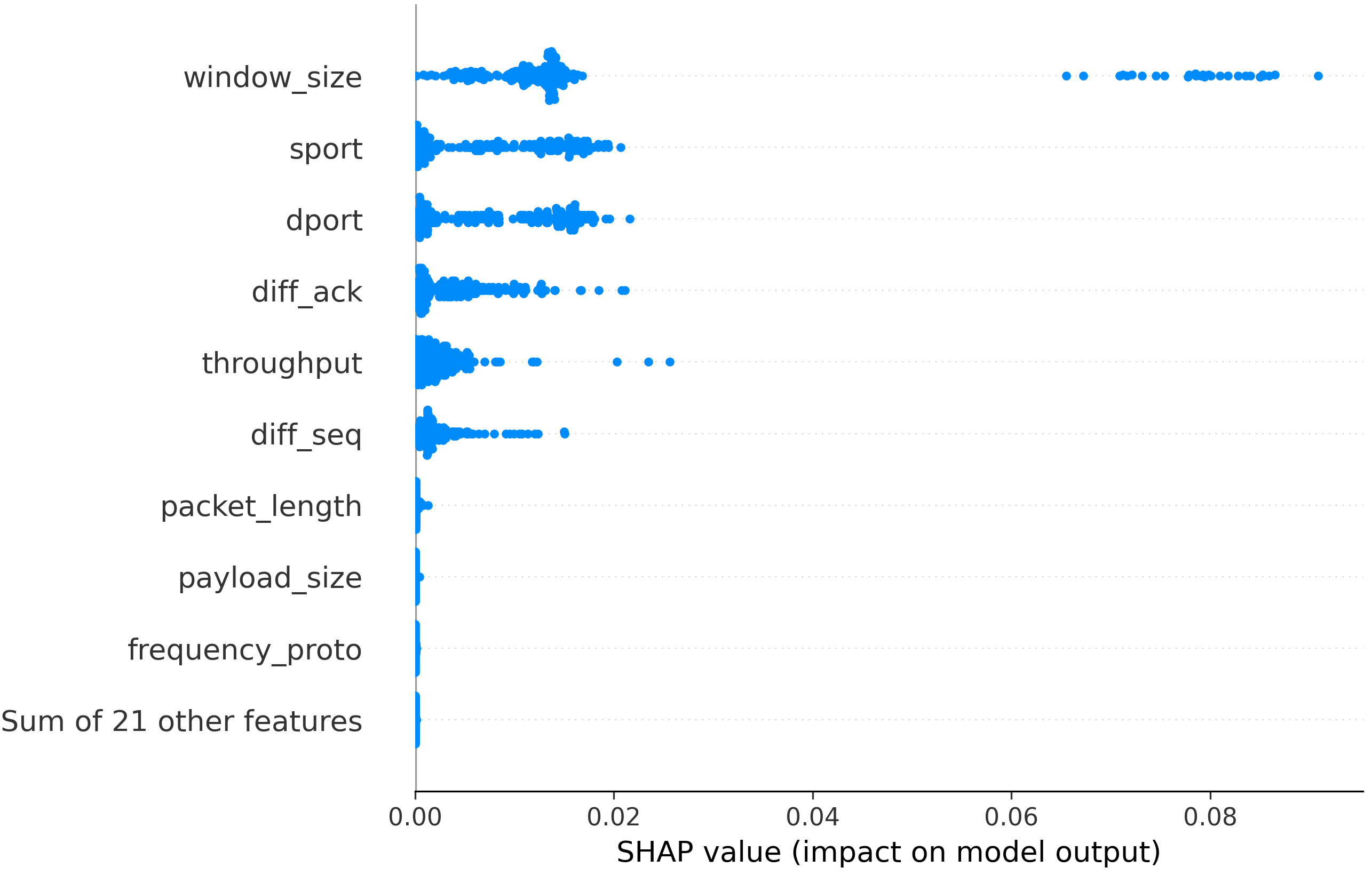}
    \caption{Feature analysis with a focus on \ac{it} layer-specific attributes, representing the \acs{dos} attack}
    \label{fig:classification-it-10}
    \vspace{-1em}
\end{figure}

\begin{figure}[htbp]
    \centering
    \includegraphics[width=\columnwidth]{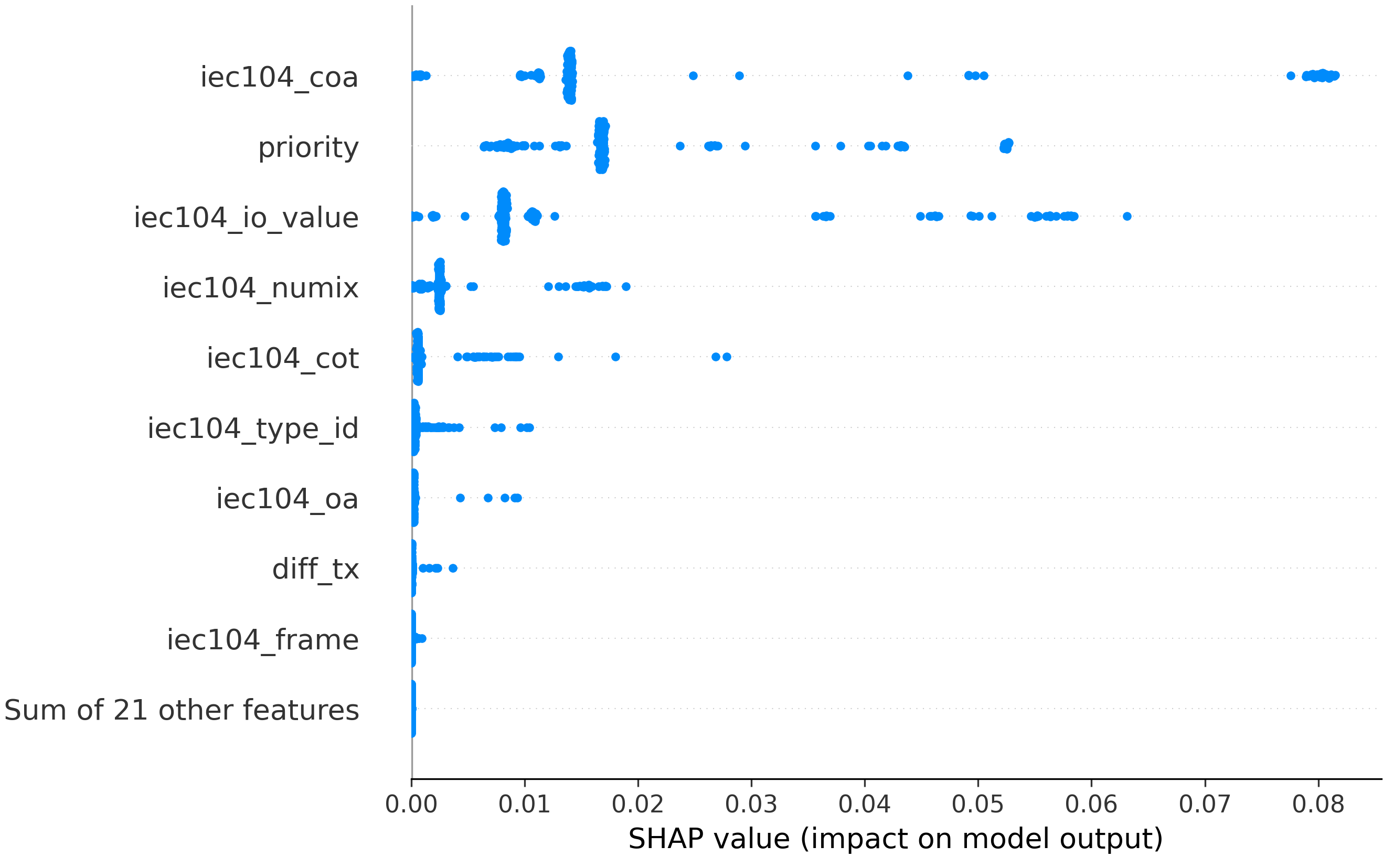}
    \caption{Feature analysis focusing on the process layer-specific attributes, specifically illustrating the \ac{dos} attack}
    \label{fig:classification-otet-10}
    \vspace{-1em}
\end{figure}

\begin{figure}[htbp]
    \centering
    \includegraphics[width=\columnwidth]{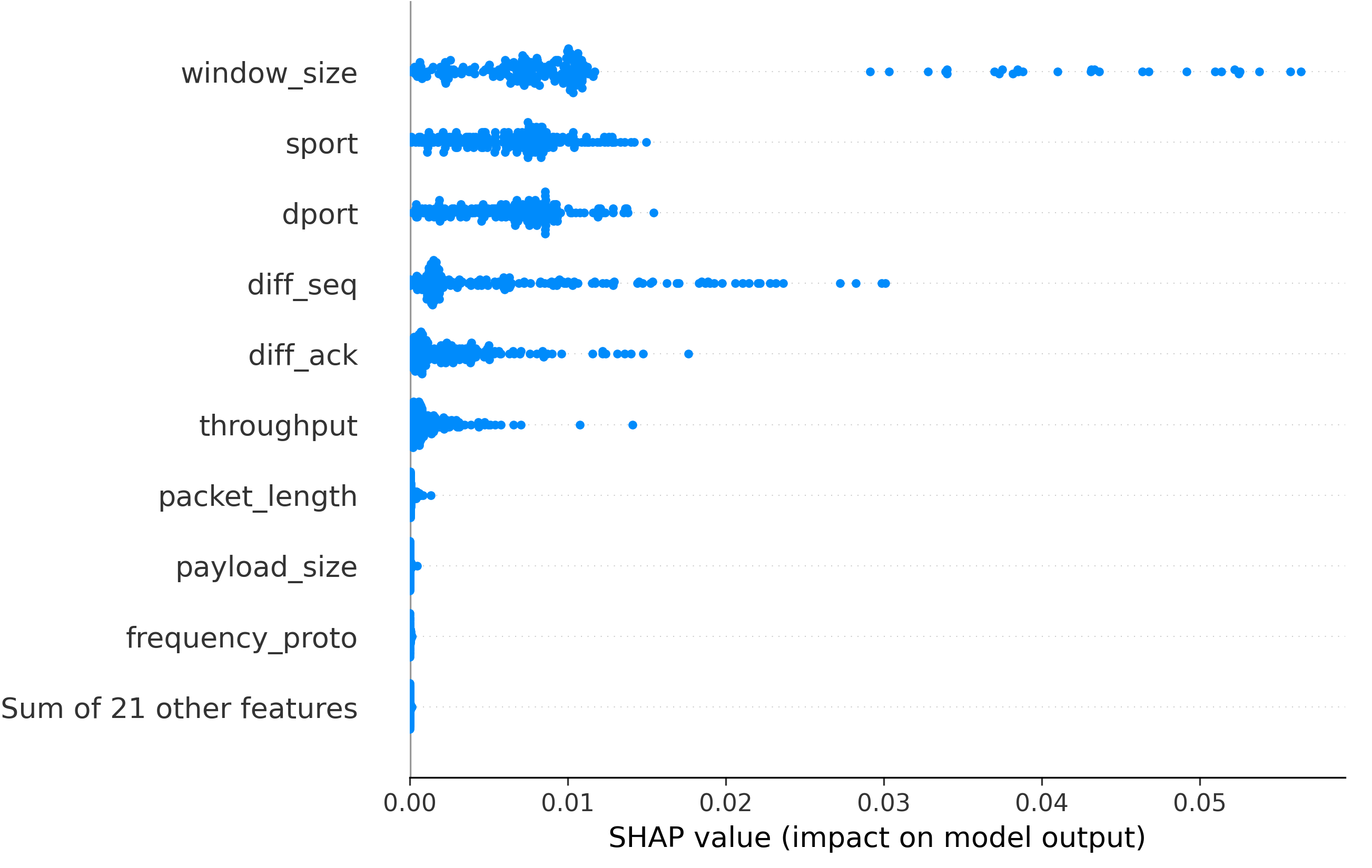}
    \caption{Feature analysis concentrating on the \ac{it} layer-specific attributes, showcasing the \ac{iec104} manipulation}
    \label{fig:classification-it-12}
    \vspace{-1em}
\end{figure}

\begin{figure}[htbp]
    \centering
    \includegraphics[width=\columnwidth]{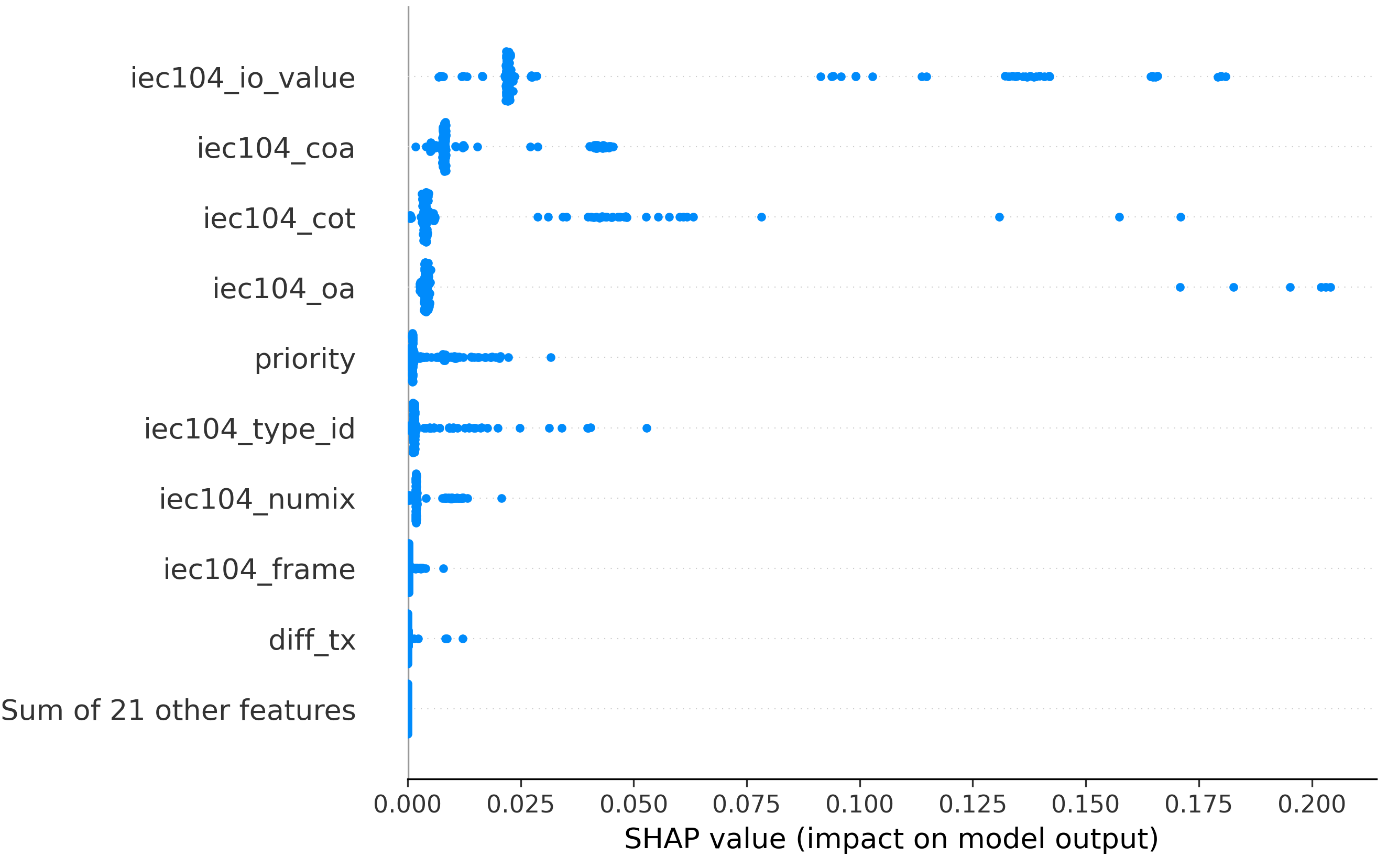}
    \caption{Feature analysis concentrating on the process layer-specific attributes, showcasing the \ac{iec104} manipulation}
    \label{fig:classification-otet-12}
    \vspace{-1em}
\end{figure}

In examining the significance of features in detection, we analyze both \ac{it} and process-focused events. Our focus includes multiple \ac{it} attacks (e.g. \ac{dos}) and \ac{iec104}-based manipulation attacks, to understand the impact of domain-specific knowledge on detection effectiveness. The observation within this investigation offer valuable insights into cyberattack detection across \ac{it}, \ac{ot}, and \ac{et} layers. 

Direct comparison of detection quality between \ac{it}-only and process-aware \ac{ids} reveals notable differences in their performance across scenarios (cf. Figure~\ref{fig:estimation_plot}). While both systems perform similarly in \ac{it}-focused events, process-aware \ac{ids} significantly excels in detecting process-focused events, particularly \ac{iec104} manipulation attacks, more accurately than \ac{it}-focused \ac{ids}.

Figure \ref{fig:classification-otet-10} depicts an attack scenario in the \ac{ot} and \ac{et} layers, highlighting general and SYN \ac{dos} attacks. However, the lack of \ac{it} data renders \ac{dos} attacks undetectable in this setup. Instead, the focus is on \ac{iec104} common object address and information object value details, inadvertently missing the real attack. Priority values are less critical here as they indicate the alert level specified by the MITRE ATT\&CK framework, crucial for real-time \ac{ids}.

Contrastingly, Figure \ref{fig:classification-it-10} presents an \ac{it}-oriented attack, including general and SYN \ac{dos}. The window size's bifurcation suggests simplicity in sent packets and limited recipient response. source ports and destination port features imply consistent host involvement, while ack and seq number variations suggest packet loss and disrupted connections in the \ac{dos} scenario.
Figure \ref{fig:classification-it-12} returns to the \ac{it} level, adding \ac{iec104} manipulations. This demonstrates the inadequacy of window size alone for attack indication, underscoring the necessity for \ac{ot} and \ac{et} specific data.
Lastly, Figure \ref{fig:classification-otet-12} explores an attack across \ac{ot} and \ac{et} layers, focusing on \ac{iec104} manipulation. The impact of IO value manipulation on system control is significant, and the recurrence of similar \ac{iec104} \acp{coas} within an attack highlights their importance. The figure also indicates a correlation of \ac{iec104} object address with value changes, often seen in attacks executed via the same spoofed line.

\subsection{Discussion}\label{sec:result_discussion}
The comparative performance analysis of \ac{it}-only and process-aware \ac{ids} demonstrates that integrating \ac{ot} and \ac{et} information significantly enhances the detection capabilities of \ac{ids} systems. This integration allows for a more nuanced detection of complex attacks, particularly those involving subtle manipulations in control systems such as \ac{iec104} value changes. The inclusion of process-specific data not only improves the accuracy of cyber threat detection but also aids in the understanding of attack dynamics across different network layers, as depicted in Figure~\ref{fig:estimation_plot}.

Both systems exhibit comparable precision, recall, and F1-score in detecting normal activities and \ac{arp} spoofing attacks, yet differ markedly in their ability to detect \ac{dos} and value manipulation attacks. The process-aware \ac{ids} outperforms the \ac{it}-only system in these scenarios, indicating its superior capability in environments prone to sophisticated cyber threats.

Overall, the findings underscore the necessity of process awareness in \ac{ids} design, particularly in environments where the integrity and stability of control systems are paramount. The advanced capabilities of process-aware \ac{ids} in detecting and responding to cyber threats affirm their critical role in safeguarding modern cyber-physical systems.

%% file: chapter5.tex
\section{Conclusion} \label{sec:conclusion}
The integration of \ac{ict} in power grids elevates both opportunities and cybersecurity risks. This paper underscores the importance of process awareness in \ac{ids}, specifically within the power grid's \ac{it}, \ac{ot}, and \ac{et} layers. We developed a prototype machine learning-based \ac{ids}, evaluated through multi-stage cyberattack simulations based on the MITRE ATT\&CK Matrix. The findings reveal that process-aware \ac{ids} significantly enhance detection capabilities for process-centric cyberattacks compared to \ac{it}-only \ac{ids}. While both types of \ac{ids} are effective for \ac{it} layer threats such as \ac{dos} and SSH brute-force attacks, process-aware systems demonstrate superior precision by utilizing domain-specific indicators. Future research should focus on advancing \ac{ids} benchmark environments and generating comprehensive datasets from co-simulations to better mirror individual infrastructure setups.